\documentclass[prl,twocolumn,aps,preprintnumbers,letterpaper,floatfix]{revtex4}
\usepackage{amsmath}
\usepackage{tipa}
\usepackage{amssymb}
\usepackage{graphicx}
\usepackage{bm}
\usepackage{enumerate}
\usepackage{subfigure}
\usepackage{color}

\newcommand{\be}{\begin{equation}}
\newcommand{\ee}{\end{equation}}
\newcommand{\bea}{\begin{eqnarray}}
\newcommand{\eea}{\end{eqnarray}}
\newcommand{\ba}{\begin{eqnarray}}
\newcommand{\ea}{\end{eqnarray}}

\newcommand{\fm}{\mathrm{fm}}

\begin{document}

\title{``Explosive regime" should dominate collisions of ultra-high energy cosmic rays}

\author{Tigran Kalaydzhyan and Edward Shuryak}
\affiliation{Department of Physics and Astronomy, Stony Brook University,\\ Stony Brook, New York 11794-3800, USA}

\date{\today}

\begin{abstract}
Since the launch of LHC experiments it has been discovered that the high multiplicity trigger in $pp, pA$ collisions  finds  events behaving differently from the typical (minimally biased) ones. In central $pPb$ case it has been proven that those possess collective phenomena known as the radial, elliptic and triangular flows, similar to what is known in heavy ion ($AA$) collisions. In this paper we argue that at the ultra-high energies, $E_{\mathrm{lab}}\sim 10^{20}\, \mathrm{eV}$, of the observed cosmic rays this regime changes from a small-probability fluctuation to a dominant one. We estimate velocity of the transverse collective expansion for the light-light and  heavy-light collisions, and find it comparable to what is observed at LHC for the central $PbPb$ case. We argue that significant changes of spectra of various secondaries associated with this phenomenon should be important for the development of the cosmic ray cascades.
 \end{abstract}

\maketitle
\section{Introduction}

Due to the air composition, passage of the utra-high energy cosmic rays through atmosphere serves as a natural nuclear collision experiment. 
By ``explosive"  we mean a dynamical regime in which
the system size becomes large comparing to the mean free path, resulting in collective flows, similar to the ones in accelerator experiments. 
 ``Ultra-high" collision energies refer to the highest ones observed in cosmic rays,
\begin{align}
E_{\mathrm{lab}} \lesssim  E_{\mathrm{max}} \sim 10^{20} \mathrm{eV} \,.
\end{align}
Detectors such as used by Pierre Auger Observatory (for recent updates see, e.g., \cite{Auger})  observe events with the energy up to
the so-called Greisen-Zatsepin-Kuzmin (GZK) bound \cite{GZK}.  For future comparison with the LHC observation it is convenient to
 convert the laboratory energy into the energy in the center of mass frame and use a standard Mandelstam invariant, assuming it is a $pp$ collision,
 \begin{align} \sqrt{s_{\mathrm{max}}}=(2E_{\mathrm{max}}m_p)^{1/2}\approx 450 \, \mathrm{TeV}\,. \end{align}
 While significantly higher than current LHC $pp$ energy $\sqrt{s_{LHC}}=8$ TeV,   the jump to it from LHC is
 comparable to that from Tevatron $\sqrt{s}=1$ TeV or  RHIC $\sqrt{s_{RHIC}}=0.5$ TeV.  In view of smooth small-power
 $s$-dependence of many observables, the extrapolation to LHC worked relatively well, and further extrapolation
  may seem to be a rather straightforward task.
 And yet, smooth extrapolations using standard event generators plus, of course, the cascade codes do not reproduce
 correctly the experimental data of the Pierre Auger collaboration (e.g., the muon size \cite{Auger}).

 This calls for a new physics: let us mention one example.
 Farrar and Allen \cite{Farrar:2013sfa} proposed a ``toy model" for the explanation of the data called ``Chiral Symmetry Restoration'' (CSR):
 their main idea is that somehow the pion production becomes suppressed. According to their simulations, a model in which mostly nucleons are produced explains the Pierre Auger data better.

   We agree that for the high density of final particles per unit rapidity, $dN/dy$,  expected at ultra-high collisions,
 corresponds to  the quark-gluon plasma (QGP) production.  QGP is indeed in a chirally restored phase with $T > T_c$.  However, as we know from the heavy ion, $pp$ and $pA$ physics at LHC, this high density matter tends to explode. The process
 of particle production ends at a certain chemical freezeout
 temperature $T_{ch}$ in a hadronic phase, close to the critical QCD temperature
 $|T_c-T_{ch}|\ll T_c$. The composition of secondaries  remains  remarkably energy- and system-independent,
 from few GeV to few TeV range of $\sqrt{s}$. There is no reason to think it will not be like that also
 in the case of cosmic rays.

  \begin{figure}[t!]
\begin{center}
\includegraphics[width=7cm]{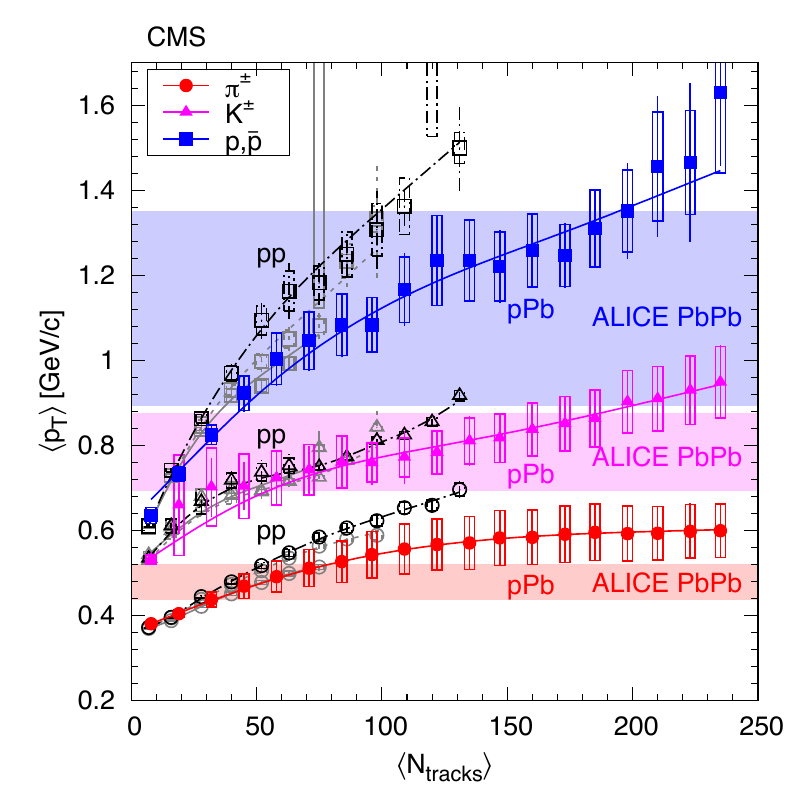}
\caption{(color online)  (From
\cite{Chatrchyan:2013eya}.) Average transverse momentum of identified charged hadrons (pions, kaons, protons)  in the range $|\eta | < 1$ as a function of the corrected track multiplicity in the rapidity acceptance $|\eta| < 2.4$, for $pp$ collisions (open symbols) at several energies, and for $pPb$ collisions (filled symbols) at $\sqrt{s_{NN}} = 5.02$~TeV. The shaded areas are the range of values for $PbPb$ collisions.
}
\label{fig_pt}
\end{center}
\end{figure}

  \begin{figure*}[t!]
\begin{center}
\subfigure[\label{curves1:a}]{\includegraphics[width=6.5cm]{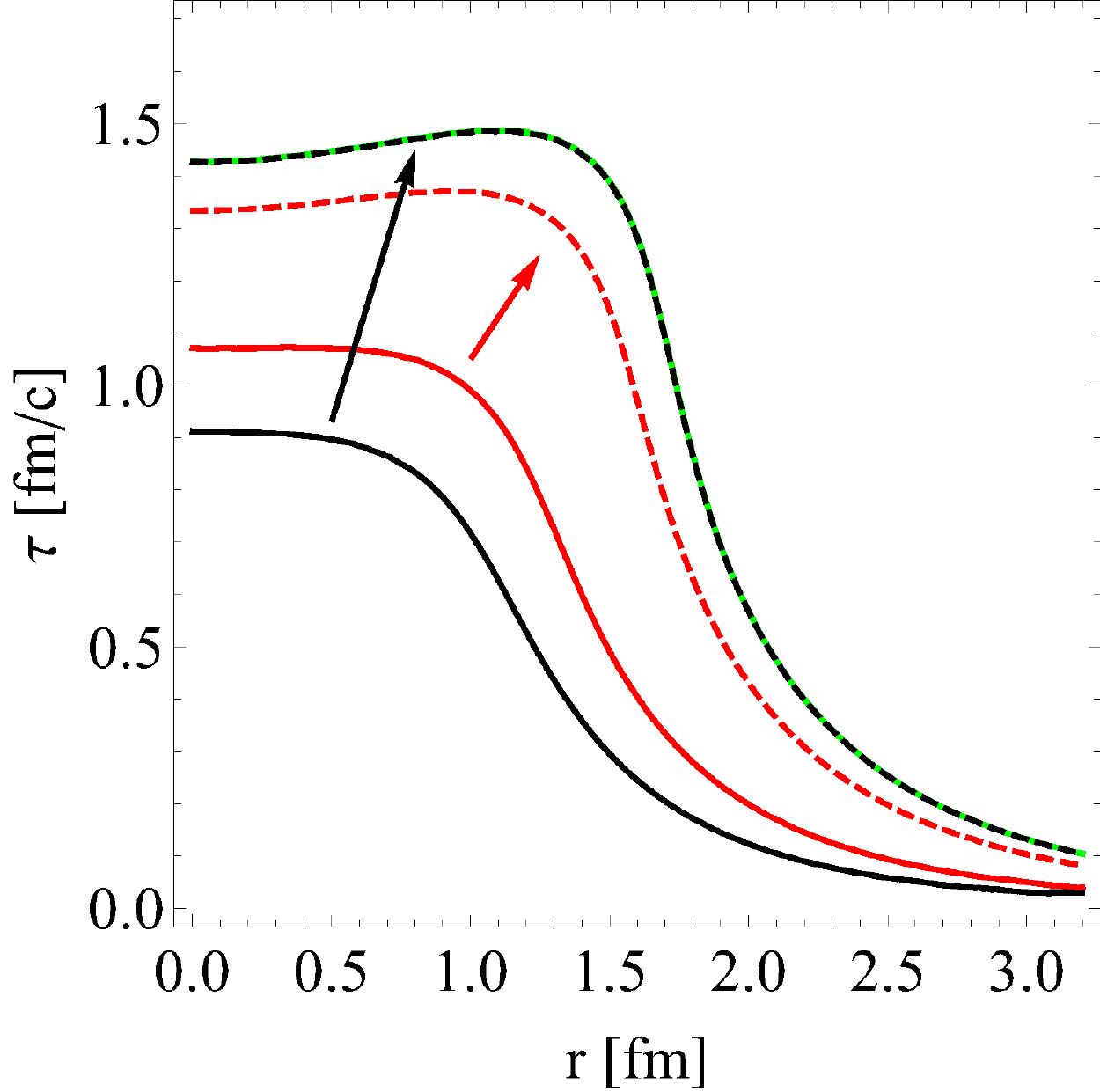}}\hspace{1cm}
\subfigure[\label{curves2:a}]{\includegraphics[width=6.3cm]{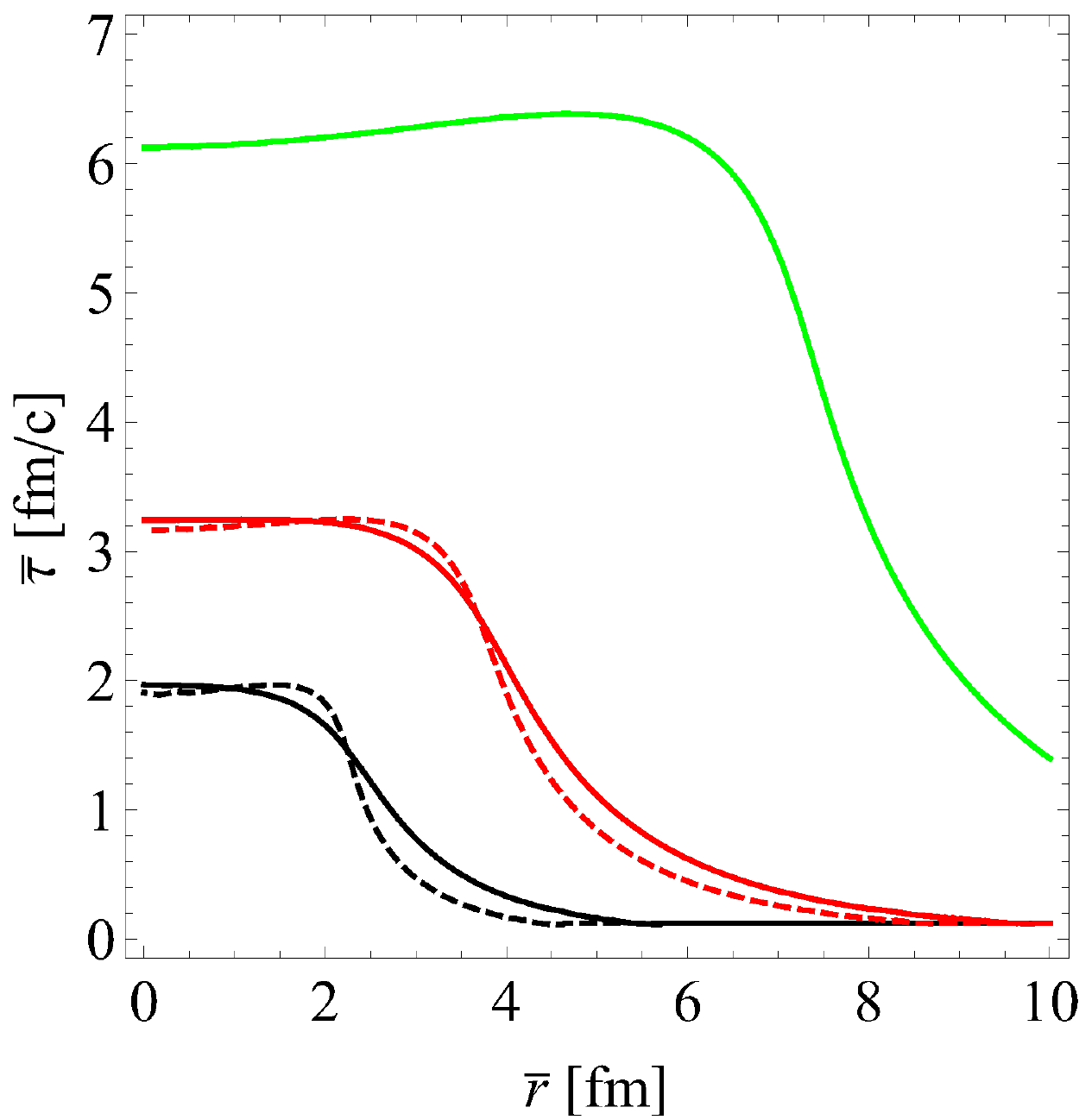}}\\
\subfigure[\label{curves1:b}]{\includegraphics[width=6.5cm]{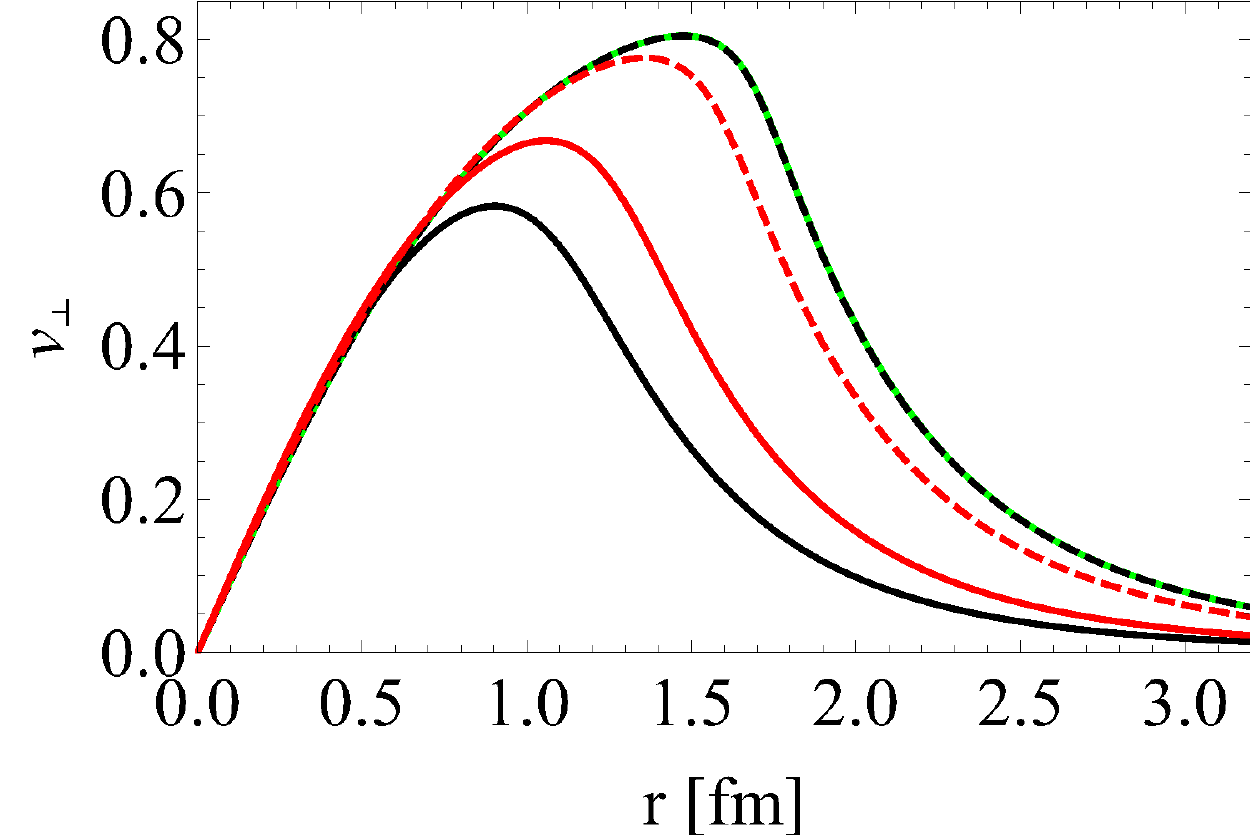}}\hspace{1cm}
\subfigure[\label{curves2:b}]{\includegraphics[width=6.5cm]{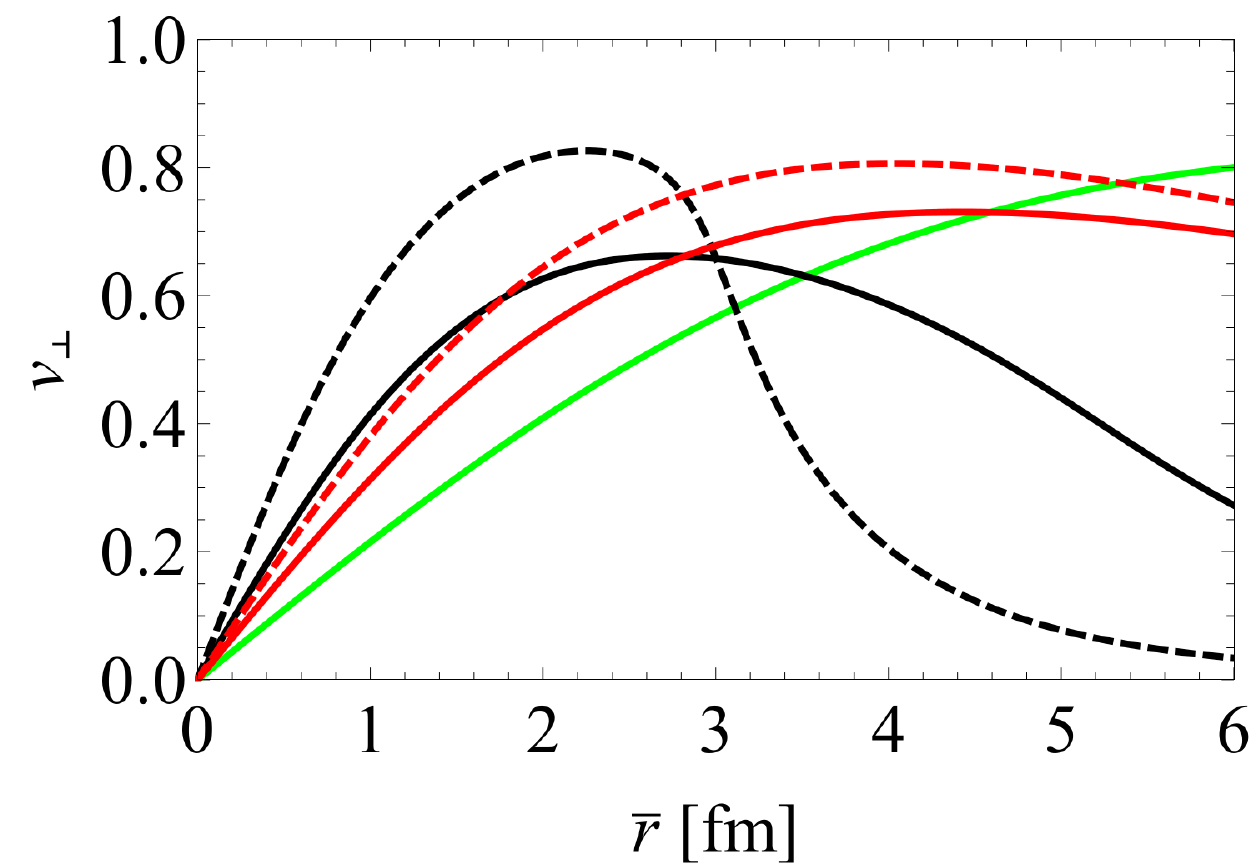}}
\caption{(color online) (a) The freezeout surfaces in the $(\tau, r)$ plane and (c) the distribution of the transverse flow velocity
on those surfaces. In both plots the green solid curve at the top is our ``benchmark", the central $PbPb$ collisions at LHC. Black solid line is for light-light
collisions, black dashed (coincident with green by chance) are light-light collisions with the size compression.  Similarly, red solid and red dashed are heavy-light collisions
without and with the size compression, respectively. Two arrows connecting the pairs are to guide the eye.
(b) The freezeout surfaces in the $(\bar\tau, \bar r)$ plane (no rescaling) and (d) the distribution of the transverse flow velocity on those surfaces. 
}
\label{fig_freezeouts}
\end{center}
\end{figure*}

Farrar and Allen argued that QGP explosion cannot be very important as it only produces elliptic flow, which is a relatively small effect --
 deformation of the angular distribution by few percents.
 While the statement itself is true, we disagree with the conclusion: the main effect of the explosion is not the elliptic
 deformation but the $radial$ flow.
It changes significantly the transverse momenta of the secondaries, especially nucleons, and therefore changes
the  production angles. Even though we do not perform any simulations of the secondary cascade in this paper and claim
no explanation of ultra-high energy data,
it is clear that the effects we discuss should change the visible size of the shower core, one of the key observables
of the cosmic ray detectors. The modifications we discuss are based on a reasonably well understood physics, and
thus should take place.

  \section{The LHC findigs }

   Searching for the quark-gluon plasma is the mainstream of the heavy ion physics. Starting from RHIC AA ($AuAu$) collisions it has been found that those are
   in a ``macroscopic" or ``explosive" regime. The spectra, two- and multi-particle correlations are well understood in
   the framework of relativistic hydrodynamics. Produced QGP has a nearly conformal equation
  of state ($\epsilon \sim T^4$, etc) and is strongly coupled, as is indicated by the small value of its viscosity. The same regime continues at the LHC domain, with even stronger collective flows.

 Collisions involving proton beams -- $pp$ and $pA$ -- are different: their spectra do not show collective flows and
 are well reproduced by models based on the independent production and breaking of the QCD strings. However,
 from the start of the LHC experiments, the high multiplicity trigger of the CMS detector was able to select events, in which the so-called ``ridge"
 phenomenon appeared \cite{Khachatryan:2010gv}. It is an  azimuthal correlation between secondaries which has a very  long range in rapidity.
 Unfortunately, those events have a very small probability, $P_{\mathrm{ridge}}^{pp}\sim 10^{-6}$, complicating so far their detailed studies.

 Subsequent studies of $pPb$ at LHC and $dAu$ at RHIC have further revealed similar correlations
 which appear with probability $P^{pA}_{\mathrm{ridge}}\sim \mathrm{few}\, \%$, for ``central" collisions. Those are studied and
 a number of observations shows that those correlations are indeed due to collective flows, similar to those seen in AA collisions \cite{Chatrchyan:2013eya}.
 Let us mention just two: (i) correlations involving 2, 4, 6 and even 8 particles indicate the same ellipticity $v_2$;
(ii) the spectra of identified secondaries show clear signatures of the radial flow.

  The simplest sign of the radial flow is that the mean transverse momentum grows with multiplicity
quite substantially, and as shown in  Fig.~\ref{fig_pt}.  The effect in $pp$ seems to be stronger than in
central $pA$, which is in turn stronger than in central $AA$ (the upper edge of the colored bands).
Even better indicator of the radial flow is the so-called $m_T$ slope of spectra of the identified secondaries,
for discussion of those and comparison to hydrodynamics see Ref.~\cite{SZ1}.

 \section{ The ``explosive regime" in ultra-high energy collisions}

  Our main statement is that at the ultra-high energies  $\sqrt{s_{\mathrm{max}}}$ observed in cosmic rays, the
   ``explosive regime"  even in $pA$ collision is expected to change from a very improbable $P\sim 10^{-6}$ fluctuation to
   the mean behavior, with $P=\mathcal{O}(1)$.  The reason for it is
   simply an increase in mean particle (entropy) density with energy $\sqrt{s}$.
   The density is the number of particles per volume, and we will evaluate both subsequently.

    The multiplicity  (per unit rapidity -- the length of the rapidity range is irrelevant as particles with very different rapidities do not interact) of $pp$ collisions and $AA$ collisions grow with energy in a bit different way,
    the fits including LHC data suggest
 \begin{align} 
 {dN_{pp} \over dy} \sim s^{0.11} , \qquad {dN_{AA} \over dy} \sim s^{0.15}\,.    
 \end{align}
(The growth is initiated by pQCD and Pomeron effects, slowly increasing the number of color exchanges and the
 number of partons/strings involved.
 Lublinsky and one of us \cite{Lublinsky:2011cw} have argued that the small extra growth in $AA$ case comes from an
 extra entropy produced by the viscous effects during the hydro evolution.)

 From the former extrapolation one gets  the enhancement factor $(s_{\mathrm{GZK}}/s_{\mathrm{LHC}})^{0.11}\approx 2.5$,
 and from the latter the corresponding factor is  $\approx 3.4$.
Since the $pA$ collision we expect to be somewhat
 in between these two regimes, we will use the following $dN/dy$ enhancement factor,
 \begin{align}
{dN^{GZK}/dy \over  dN^{LHC}/dy}\approx 3 \,,
  \end{align}
from the LHC to the ultra-high energy edge for all types of collisions.  From Fig.~\ref{fig_pt} alone one can thus expect certain growth of the mean $p_T$.

Unfortunately, it is not very clear what is the characteristic physical size of the system produced in high multiplicity $pp$ collisions
at the LHC: this issue is model-dependent and is intensively discussed at the moment.

(It can be deduced evolving hydrodynamics backwards in time -- from an observed final state to the initial size. However, as we already mentioned, in the $pp$ case no convincing
evidences for collective flows were obtained so far due to a small probabilty/statistics.)

 The problem of the initial system size
 becomes more clear for the collisions involving nuclei.
In fact,  primary collisions and subsequent cascade of ultra high energy cosmic rays all happen in the Earth atmosphere,
 so the targets are not protons but light $N$ or $O$ nuclei. Furthermore, the projectiles themselves are also most likely
 to be nuclei:  the distribution of primary collisions is incompatible with protons. It is
  either also some light nuclei or even a mixture including heavier ones, believed to be up to $Fe$ \cite{Auger}.

 Taking into account large $pp$ cross section at ultra high energies, $\sim 150\, \mathrm{mb}$, one finds that its typical impact parameters
 $b\approx 2 \, \fm$. Thus the range of the interaction in the transverse plane is comparable to the radius of the light nuclei (oxygen $R_O\approx 3\, \fm$) and therefore even in the $pO$ collisions
most of its 16 nucleons would become collision ``participants". For light-light AA collisions like $OO$ the number of participants changes from 32
 (central) to zero. Accidentally, the average number of participants is comparable to the
average  number of participant nucleons $\langle N_p \rangle\approx 16$
  in central $pPb$ collisions at the LHC.
  So, in the light-light category we assume the initial transverse size
  to be  $R_O$ and multiplicity to 3 times that in central $pPb$ collisions at the LHC.
 For heavy-light collisions (like $FeO$)  it is likely that most nucleons participate, or $N_p^{FeO} \sim 70$.  The size we assume to be
$R_{Fe}\approx 4.8\, \fm $. The multiplicity scales as the number of participants, namely heavy-light is $70/16$ of that in the light-light category.

\begin{figure*}[t!]
\begin{center}
\subfigure[\label{curves1:a}]{\includegraphics[width=7cm]{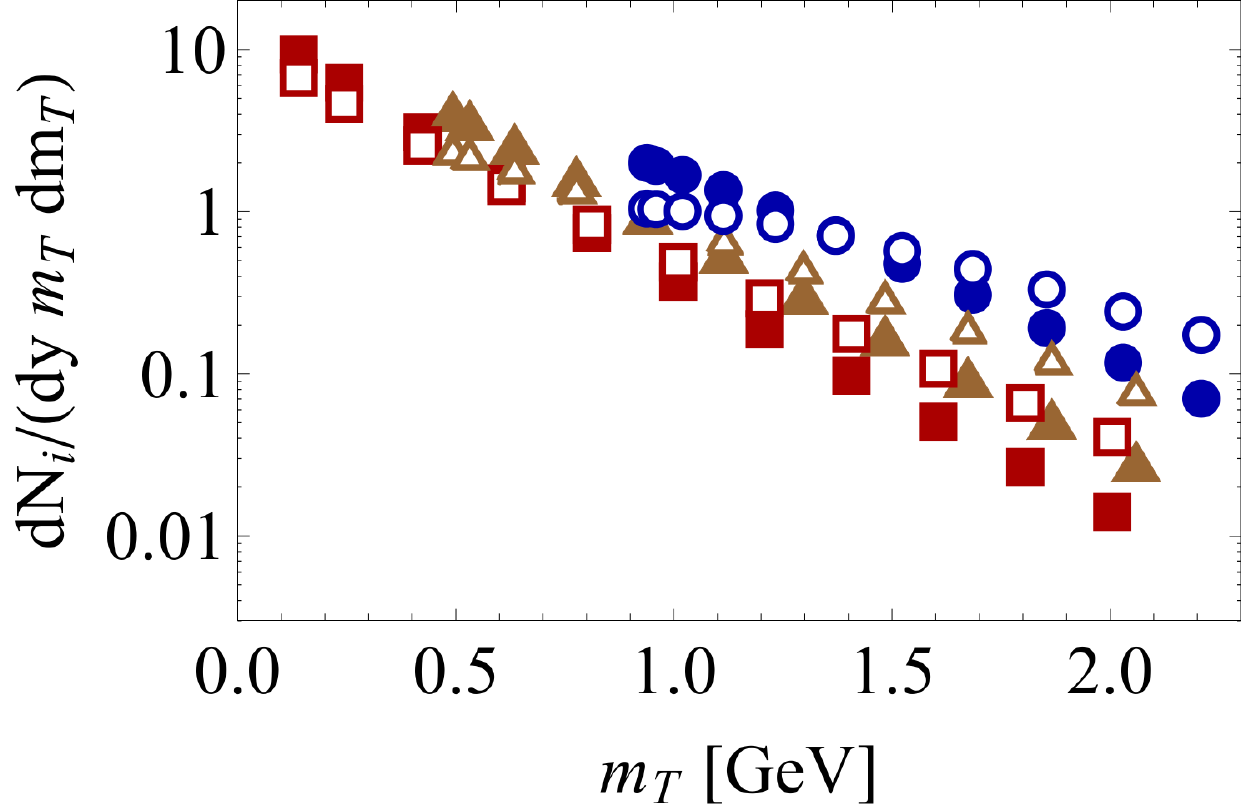}}\hspace{1cm}
\subfigure[\label{curves1:b}]{\includegraphics[width=7cm]{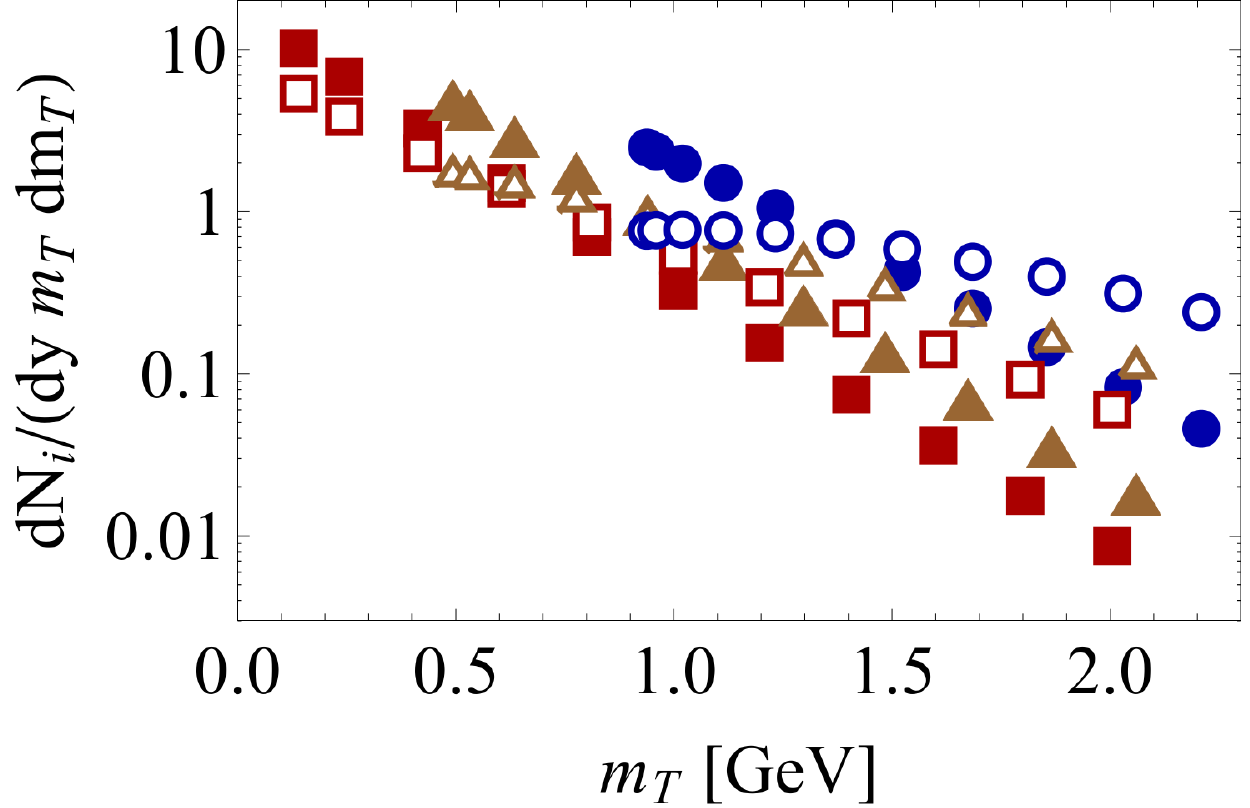}}
\caption{(color online) Normalized spectra of pions (squares), kaons (triangles) and protons (discs) for the (a) heavy-light (e.g. $FeO$) and (b) light-light (e.g. $pO$) collisions. Open symbols correspond to the ``compressed" cases, explained in the text.}
\label{fig_spectra}
\end{center}
\end{figure*}

\begin{table}[t!]
\centering
\begin{tabular}{| c || c | c | c | c | c |}
\hline
particles &  \,\,\,FeO\,\,\, & \,FeO comp.\, & \,\,\,pO\,\,\, & \,pO comp.\, & \,\,PbPb\,\,\\ \hline \hline
$\pi^\pm$   	&  0.56    &  0.69    &  0.53    &  0.76    &  0.73        \\\hline
$K^\pm$ 	    &  0.71    &  0.88    &  0.66    &  0.96    &  0.92        \\\hline
$p,\, \bar p$ 	&  0.90    &  1.09    &  0.83    &  1.17    &  1.13         \\ \hline
\end{tabular}
\caption{Mean $p_T$ [GeV/c] for pions, kaons and protons obtained from the particle spectra.
By ``comp" we mean compressed initial state, as explained in the text.
}\label{pttable}
\end{table}

Hydrodynamics  provides a connection between the initial and the final properties of the system, and one should eventually find out the initial size, assuming the hydro works.
For the problem at hand -- to see how
the result depends on the size of the system  -- it is convenient to follow the paper of Zahed and one of us \cite{SZ1}, in which the   radial flow is studied with the use of the (azimuthal angle and rapidity independent) Gubser's solution \cite{Gubser:2010ze}.
From the proper time and transverse radius
  $\bar{\tau},\bar{r}$ (with the bar) we proceed 
to dimensionless variables
$  \tau=q \bar{\tau}, \,\,\,r=q \bar{r} $
by the scaling factor $q$, the first parameter of the model. The factor $q$ is an inverse characteristic transverse size of the system at the beginning of the hydro phase (taken from e.g. the nuclei radii, $pp$ total cross section, etc.). The
solution
 for the transverse velocity and the energy density reads
\begin{align}
 & v_\perp(\tau, r)=  \frac{2 \tau r }{1 + \tau^2 + r^2}\,, \\
 & {\epsilon \over q^4}  =  \frac{\hat{\epsilon}_0\, 2^{8/3}}{\tau^{4/3}\left[1+2(\tau^2 +
r^2)+(\tau^2-r^2)^2\right]^{4/3}}\,.
\end{align}
 The energy density
has a second dimensionless
parameter  $\hat{\epsilon}_0 $
related to the multiplicity $dN/dy$,
\be \hat{\epsilon}_0=f_*^{-1/3} \left({3 \over 16 \pi} {dS \over d\eta}\right)^{4/3}, \ee
where $f_*=11$ is the number of effective degrees of freedom in QGP and the entropy per (pseudo)rapidity, $dS /d\eta\approx 7.5\, dN_{ch}/d\eta$, is proportional to the number of charged particles per unit rapidity \cite{Gubser:2010ze} .
Putting those to the same hydro solution we find the solid freezeout curves shown in Fig.~\ref{fig_freezeouts}.
The left hand side of the plots shows them in rescaled coordinates: two coinciding curves correspond to
the same flow, as the traverse velocity depends only on $\tau, r$. The right hand side shows absolute coordinates, in $\fm$:
in this case one can better compare the shapes of the surfaces for two initial size options we use.

We now remind that similar ``naive" estimates would predict that the radial flow for $pPb$ at the LHC is $weaker$ than in central AA (the benchmark). As we already discussed in the Introduction, this contradicts the observations \cite{Chatrchyan:2013eya}, and the  solution proposed in \cite{Kalaydzhyan:2014zqa} is the so-called ``spaghetti collapse" (strings stretched between participating nuclei are attracted to each other and lead to a compression of the system). As a possibility, we would like to include this phenomenon as well. We do so in a very schematic way, by reducing the initial radius of the system by $\Delta R=- 1\, \fm$. (The magnitude of the compression cannot be larger because at time exceeding 1 fm/c strings breaking occurs.)

The results are shown in Fig.~\ref{fig_freezeouts} by the dashed lines. As one can see, the compression increases the flow.
One can use the freezeout curves and substitute them to the Cooper-Frye formula \cite{Cooper:1974mv} in order to obtain $m_T$ particle spectra, where $m_T \equiv \sqrt{p_T^2 + m^2}$. As one can see from the Fig.~\ref{fig_spectra}, the compression affects the spectra, especially of the secondary protons. The $m_T$-scaling (exponential distributions with the same slope for all particles) is violated in the compressed case (open symbols) stronger, which is a clear evidence of the collective behavior. At the LHC such violation becomes significant only for rare high-multiplicity events \cite{Chatrchyan:2013eya}, while here it becomes visible for average multiplicity. The mean $p_T$ are also calculated from the spectra and presented in the Table~\ref{pttable}. The $\langle p_T \rangle$ is increased further if one takes into account the compression of the system -- for the $pO$ collision the effect is especially pronounced and leads to a strong enhancement.

\section{ Summary and discussion}
We argue that the ``explosive" regime, seen in $central$ $pA$ collisions at the LHC with few percent probability (and, of course, in $AA$)
 should become dominant (with probability $\mathcal{O}(1)$) in ultra-high energy $pA$ collisions. It is even more clear for collisions of light-light ($N, O$) or heavy-light $(FeO)$ nuclei.

 We performed some estimates of the magnitude of the collective flow for those cases, and conclude that -- within the uncertainties related to an equilibration mechanism at early stages -- it is likely to be quite similar to those in the central $PbPb$ collisions at LHC.

 Needless to say, we only did analytic estimates: predictions can obviously be made more accurate with
 more efforts. Furthermore, it would be needed to use
   the corresponding spectra as an input for the cosmic ray cascades, instead of the (extrapolated) $pp$ spectra, to see how much the actual observables will be changed.

{\bf Acknowledgements.}
This work was supported in part by the U.S. Department of Energy under Contract No. DE-FG-88ER40388.

\end{document}